\documentclass[12pt,elsart,epsfig]{article}     
\usepackage{epsfig}

\begin {document}
\vskip 15mm
\begin {center}
{\Large \bf Partial wave analysis of $\bar pp$ annihilation channels
in flight 
\vskip 1mm
with $I = 1, C = +1$}
\vskip 5 mm

{A.V. Anisovich$^c$, C.A. Baker$^b$, C.J. Batty$^b$, D.V. Bugg$^a$,
V.A. Nikonov$^c$, A.V. Sarantsev$^c$, V.V. Sarantsev$^c$, 
B.S.~Zou$^{a}$ \footnote{Now at IHEP, Beijinj 100039, China} \\
{\normalsize $^a$ \it Queen Mary and Westfield College, London E1\,4NS, UK}\\
{\normalsize $^b$ \it Rutherford Appleton Laboratory, Chilton, Didcot OX11 0QX,UK}\\
{\normalsize $^c$ \it St. Petersburg Nuclear Physics Institute, Gatchina, 
St. Petersburg district, 188350, Russia}\\ 
[3mm]}
\end {center}

\begin{abstract}
A combined analysis is reported of $3\pi ^0$, $\pi ^0\eta$ and $\pi ^0 \eta '$
data in the mass range 1960 to 2410 MeV.
This analysis is made consistent also with $\eta \eta \pi ^0$ data,
reported separately.
The analysis requires $s$-channel resonances with a spectrum close
to that published earlier for $C = +1$ states with $I = 0$; masses for $I=1$
states are
lower on average by 20 MeV.
Two alternative solutions
are found, differing only for $J^P = 2^+ $ and $4^+$ states by small amounts
in masses and widths.
Both $3\pi ^0$ and $\eta \pi ^0$ data prefer one of these two solutions.
For this preferred solution, observed states have $J^{PC}$, masses and
widths ($M$, $\Gamma $) in MeV as follows:
$4^{-+}$: ($2250 \pm 15$, $215 \pm 25)$,
$4^{++}$: ($2255 \pm 40$, $330 ^{+110}_{-50}$) and
($2005 ^{+25}_{-45}$, $180 \pm 30$),
$3^{++}$: ($2275 \pm 35$, $350 ^{+100}_{-50}$) and
($2031 \pm 12$, $150 \pm 18$),
$2^{-+}$: ($2245 \pm 60$, $320 ^{+100}_{-40})$ and
($2005 \pm 15$, $200 \pm 40)$,
$2^{++}$: ($2255 \pm 20$, $230 \pm 15$), ($2175 \pm 40$,
$310^{+90}_{-45})$ and ($2030 \pm 20$, $205 \pm 30$),
and $1^{++}$: ($2270 ^{+55}_{-40}$, $305 ^{+70}_{-35})$.
There are indications of further $2^{-+}$, $2^{++}$ and $1^{++}$
contributions just below the available mass range, and also a
$0^{++}$ state at $\sim 2025$ MeV.

\end{abstract}
\vskip 4mm

Data for $\bar pp \to 3\pi ^0$ have been reported earlier [1] from the
Crystal Barrel experiment at LEAR in the momentum range 600 to 1940 MeV/c.
Data from channels $\bar pp \to \pi ^0 \eta$ and $\pi ^0 \eta '$ have also
been presented [2].
There is evidence for a number of $s$-channel resonances with
similar masses and widths in the two analyses.
The objective here is to report a combined analysis with consistent
resonance parameters in all three sets of data.

A related analysis of $\bar pp \to \eta \eta \pi ^0$  in reported
separately [3].
Those data provide evidence for two $0^-$ resonances which are less
conspicuous in the $3\pi ^0$ data discussed here.
The present analysis uses the parameters of those resonances.
Conversely, the analysis of $\eta \eta \pi ^0$ uses parameters of
resonances reported here.
It is useful to examine the sensitivity
of each set of data to individual resonances.

A second objective is to compare resonance masses and widths with a combined
analysis reported earlier [4] of $I = 0$, $C = +1$ channels
$\pi ^0 \pi ^0$, $\eta \eta$, $\eta \eta '$, $\eta \pi ^0 \pi ^0$ and
$\pi ^+\pi ^-$.
There, a complete spectrum of the $q\bar q$ states expected in this
mass range was found, as well as some additional states.
If mass differences between $u$ and $d$ quarks are small, as is generally
believed,
the spectra for $I = 1$ and 0 should be similar.
This is what we find.

We outline first the considerations going into the partial wave analysis.
The earlier study of $3\pi ^0$ data fitted magnitudes and phases of
amplitudes separately to data at individual momenta.
Those results were then interpreted in terms of $s$-channel resonances
for $J^P = 4^+$, $3^+$, $2^+$ and $1^+$.
For $\pi ^0 \eta$ and $\pi ^0 \eta '$,
magnitudes and phases of $\pi ^0\eta$ and $\pi ^0 \eta '$ amplitudes
were close to SU(3) relations [2].
The composition of $\eta$ and $\eta '$ are well known from a study of many
branching ratios [5] to be
\begin {eqnarray}
|\eta > = \cos \theta |\frac {u\bar u + d\bar d}{\sqrt {2}}> -
          \sin \theta |s\bar s > \\
|\eta '>= \sin \theta |\frac {u\bar u + d\bar d}{\sqrt {2}}> +
          \cos \theta |s\bar s >,
\end {eqnarray}
with $\cos \theta \simeq 0.8$ and $\sin \theta \simeq 0.6$.
The same SU(3) constraints are  applied here.

Partial wave amplitudes are expressed as sums of $s$-channel resonances
plus backgrounds. Each resonance is fitted to a
Breit-Wigner amplitude of constant width with real coupling constant $g$ and
phase angle $\phi$:
\begin {equation}
f = \frac {\sqrt {\rho _{\bar pp}(p)}}{k}\frac {g \exp (i\phi )
B(p)B(q)}{M^2 - s - iM\Gamma }.
\end {equation}
Backgrounds are parametrised as constants or linear
functions of $s$, or as resonances below the $\bar pp$ threshold.
These assumptions impose the important constraint of analyticity.
Blatt-Weisskopf centrifugal barrier factors $B$ are included explicitly for
coupling to $\bar pp$ (momentum $p$ in the overall centre of mass frame) and
coupling to the decay channel (momentum $q$);
the radius of the centrifugal barrier is set to 0.83 fm from Ref. [4].
The factor $1/k$ allows for the flux in the $\bar pp$ entrance channel,
and $\rho _{\bar pp}$ is the phase space in the $\bar pp$ channel.

Partial waves with $J^P = 2^+$ and $4^+$ may couple to $\ell = J \pm 1$ in
the $\bar pp$ channel, e.g. $^3F_2$ and $^3P_2$.
Multiple scattering through the resonance is expected to lead to
approximately the same phase $\phi$ for both $\ell$ values.
For $I = 0$ $C = +1$, phases are accurately determined; all phase differences
between $\ell = J \pm 1$ lie in the range $0 \pm 15 ^{\circ}$.
For present data, they are less accurately determined (because of the
lack of polarisation data) but are consistent with zero.
We therefore fit the ratio of coupling constants $g_{J+1}/g_{J-1}$ to
a real ratio $r_J$.
Most states turn out to be dominantly $L = J - 1$ or
$L = J + 1$, and the larger partial wave governs the determination
of resonance masses and widths.

The amplitude analysis turns out to be much less secure than for
$I = 0$, $C = +1$ for several reasons.
The fundamental reason is that no data are
available from a polarised target, as was the case for $I = 0$ in the
$\pi ^-\pi ^+$ channel.
Polarisation data play two fundamental roles.
Firstly, they separate amplitudes with helicities 0 and 1 in the
initial state, hence $\ell = J \pm 1$.
In the present analysis, separation between $^3P_2$ and $^3F_2$
amplitudes and between $^3F_4$ and $^3H_4$ is hampered by the
absence of such polarisation information.
The second role of polarisation data is that they are phase sensitive.
Polarisation measures the imaginary part of interferences between
partial waves, while differential cross sections are sensitive to
the real parts of interferences.
For $I = 0$, $C = +1$, the availability of both
differential cross sections and polarisations
puts tight constraints on all phases.
For the present $I = 1$, $C = +1$ channels, relative phases are
poorly determined when the phase angle between partial waves is close
to 0 or $\pi$, because of the lack of polarisation data.
This leads to larger errors for several resonances.

\begin{figure}
\begin{center}
\vskip -25mm
\epsfig{file=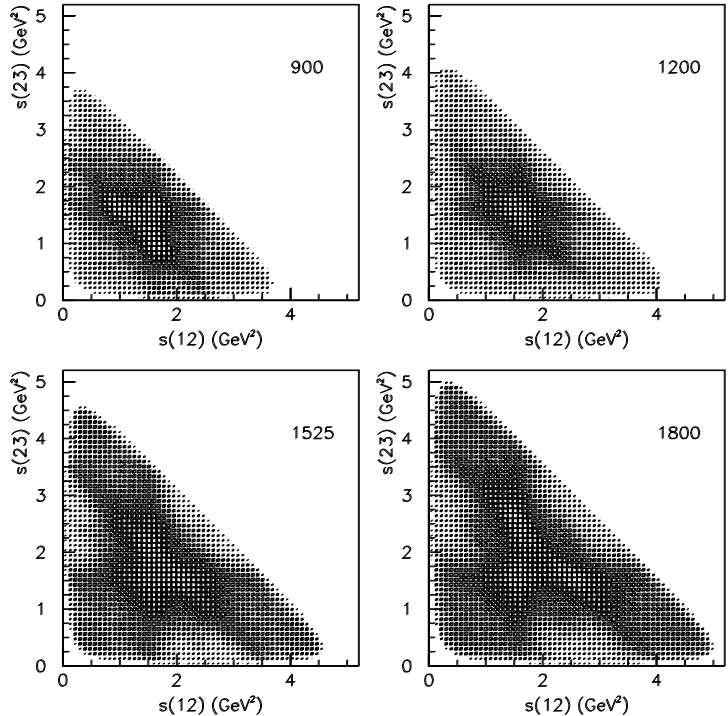,width=11cm}\
\vskip -110.45mm
\hskip 3.05mm
\epsfig{file=DALITZ.EPS,width=11cm}\
~\
\vskip -8mm
\caption{Dalitz plots for data; numbers in each panel indicate beam
momenta in MeV/c. }
\end{center}
\end{figure}

A second problem is as follows.
Dalitz plots for $3\pi ^0$ data are shown at four representative momenta
in Fig. 1.
There is just one dominant signal, $f_2(1270)\pi$, from which to
search for resonances.
Underneath the $f_2(1270)$ bands is a broad physics background.
This is fitted as $\sigma \pi$, where $\sigma$ stands for the
$\pi \pi$ S-wave amplitude as parametrised in Ref. [6].
Because of the sixfold symmetry of the Dalitz plot, it is hard to
separate contributions from low and high masses in $\pi \pi$.
The $\sigma $ amplitude varies slowly with $s$ and
its coupling constant is likely to have some $s$-dependence.
We assume the coupling constant may vary linearly with $s$.
There must also be contributions from so-called triangle diagrams [7].
One of the pions from the decay of $f_2(1270)$ may
rescatter from the spectator pion, producing a new resonance.
In principle such processes are calculable. They lead to a logarithmic
variation of the amplitude with $s$.
Such processes cannot in practice be separated from
the $\pi \pi$ S-wave amplitude.
In summary, there is an intrinsic uncertainty
about how to parametrise the background. Different parametrisations
lead to somewhat different interferences with $f_2(1270)$ bands.
Since the broad background appears dominantly in the $J^P = 0^-$ channel,
these differences introduce uncertainties mostly into the determination
of singlet partial waves with $J^P = 0^-$, $2^-$ and $4^-$.

\begin{figure}
\begin{center}
\vskip -24.0mm
\hskip -2.90mm
\epsfig{file=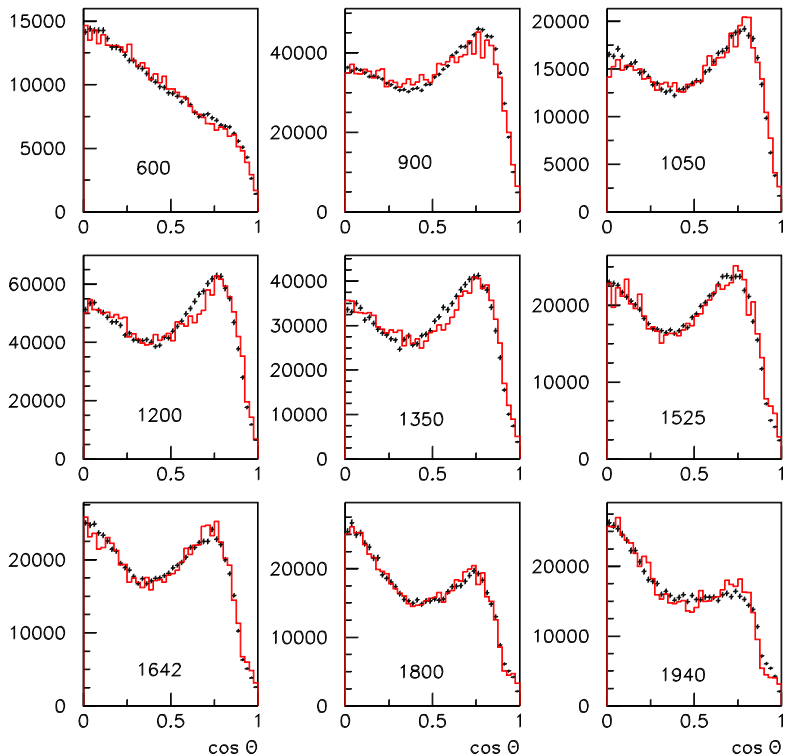,width=11.6cm}\
\vskip -116.45mm
\epsfig{file=HIST2.EPS,width=11.6cm}\
~\
\vskip -8mm
\caption{Angular distributions for production of $f_2(1270)\pi$ for
events with at least one value of $M(\pi \pi)$ in the range 1175--1375 MeV.
Histograms show the partial wave fit.
Numbers in each panel indicate the beam momentum in MeV/c.}
\end{center}
\end{figure}

The fit to angular distributions in Fig. 2 at 1200 and 1350 MeV/c
is not perfect.
The problem is associated with the crossing of three $f_2(1270)$
bands, visible on Fig. 1.
A possible explanation is that triangle effects of Ref. [7] will have
maximum effect at this triple intersection.

A third problem, purely experimental, is that the production angular
distribution for $f_2(1270)$ shows an extremely rapid change
between the two lowest momenta, 600 and 900 MeV/c. It is illustrated
in the first two panels of Fig. 2. This makes it hard to determine
resonance parameters for the lower group of $s$-channel resonances
which cluster in this range. The analysis of $I = 0$, $C = +1$ was
on firmer ground, because of the availability of
data for $\bar pp \to \pi ^- \pi ^+$ at 100 MeV/c steps from 360 to
900 MeV/c, both differential cross sections and polarisations;
those data extend down to a mass of 1910 MeV.

A defect in our earlier analyses in Refs. [1] and [2] was that
$3\pi ^0$ data were fitted with only two $2^+$ resonances, while
those for $\pi ^0 \eta$ and $\pi ^0 \eta '$
were fitted with three. They also used
different values of the ratios $r_2$ between $^3F_2$ and $^3P_2$
amplitudes and likewise for $r_4$ between $^3H_4$ and $^3F_4$.
These defects are rectified here.
It turns out to be quite difficult to find a combined fit
to all data with consistent values of $r_2$ and $r_4$.
A good fit requires that $r_2$
varies rapidly over the available mass range.
Four $2^+$ states are required.
This agrees with $I = 0$, $C = +1$,
where two dominantly $^3P_2$ states were found at 1934 and 2210 MeV and
two dominantly $^3F_2$ states at 2010 and 2293 MeV.
A similar pattern emerges here.
As a result, fits to data have changed significantly
from the earlier publications for partial waves with $J^P = 2^+$ and
$4^+$.

This change is particularly large for $3\pi ^0$ data at the lowest two
momenta.
The fit shown in Ref. [1] had a large $^3P_1$ intensity there.
However, we now find that adding further $^3P_2$ states at $\sim 1950$
and 2175 MeV produces a very large improvement in the fit,
by $\sim 10500$ in log  likelihood.
There is large cross-talk at low momenta
between $^3P_2$ and $^3P_1$ and, to a
lesser extent, with  $^3F_3$; these partial waves
all produce $f_2(1270)\pi$ amplitudes with $L = 1$ and 3 in the final state
and differ
only by Clebsch-Gordan coefficients for different helicities.
In the new fit, the $^3P_1$ amplitudes shrink to quite small values,
and $^3P_2$ and $^3F_3$ amplitudes grow to large values.
This change is a direct consequence of the simultaneous fit to three
channels of data.

We now deal with some technicalities.
Small contributions due to $f_0(980)$, $f_0(1500)$ and $f_2(1565)$ are
visible in  mass projections of Fig. 3.
The structure at $s = 2 - 2.5$ GeV$^2$ in the first panel (600 MeV/c)
is largely  due to $f_2(1565)$, not $f_0(1500)$.
This contribution dies away rapidly at higher momenta.
It is parametrised by the form given in Ref. [8].
Contributions from $f_0(1370)$ may be separated reliably
from those due to $f_2(1270)$ but are small.
The fit to $\eta \eta \pi ^0$ requires contributions also from
$f_0(1770)$, $f_2(1980)$ and $f_0(2105)$.

\begin{figure}
\begin{center}
\vskip -24.0mm
\hskip -2.9mm
\epsfig{file=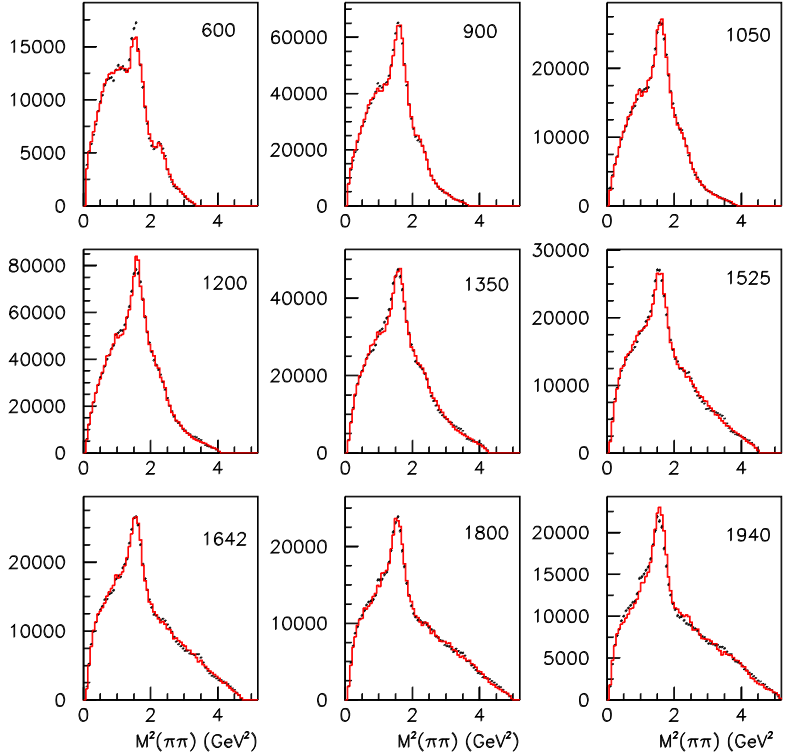,width=11.6cm}\
\vskip -116.45mm
\epsfig{file=HIST1.EPS,width=11.6cm}\
~\
\vskip -8mm
\caption{Projections on to $M^2(\eta \eta)$; histograms show the
fit. Numbers in each panel indicate beam momenta in MeV/c.}
\end{center}
\end{figure}

No significant physics can be extracted from these small amplitudes.
If they are fitted freely to all partial waves, there is
the danger that they drift to sizeable magnitudes with large destructive
interferences between them.
This is a well known form of instability.
To avoid it, a penalty function is introduced for magnitudes $\Lambda $
of amplitudes.
This penalty function takes the form of contributions to $\chi ^2$:
$$\chi _i ^2 = [(\Lambda - \Lambda _0)/\delta \Lambda ]_i^2.$$
Here $\Lambda _0$ is a target value, zero for $f_0(980)$, $f_0(1370)$,
$f_2(1565)$.
The denominator $\delta \Lambda $ is adjusted so that
the magnitude of each partial wave contributes up to $\chi ^2 = 9$
(i.e. $3\sigma$) to the penalty function.
Those amplitudes which are really needed feel little influence from the
penalty function compared with very large contributions to log
likelihood from individual data points; amplitudes which are not needed settle
close to zero. In practice this simple procedure stabilises the small
partial waves very effectively.
It contributes $\sim 650$ to $\chi ^2$ compared with
$\sim 650,000$ for log likelihood.

\begin{figure}
\begin{center}
\vskip -24mm
\hskip -2.85mm
\epsfig{file=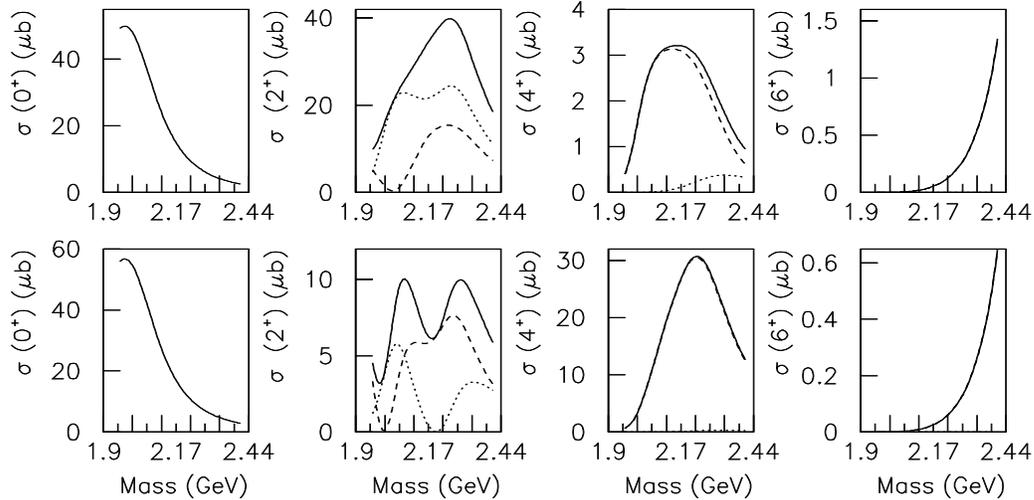,width=15cm}\
\vskip -155.45mm
\hskip 3.05cm
\epsfig{file=FPIE1.EPS,width=15cm}\
~\
\vskip -65mm
\caption{Intensities of $0^+$, $2^+$, $4^+$ and $6^+$ partial waves
in $\bar pp \to \pi ^0 \eta$ for solution 2 (upper row) and
solution 1 (lower row).
Dashed curves show intensities for $^3P_2$ and $^3F_4$, dotted curves for
$^3F_2$ and $^3H_4$, full curves their sum.}
\end {center}
\end {figure}

The $f_0(1500) \to \eta \eta $ is clearly visible in $\eta \eta \pi ^0$
data and is well determined there.
A first pass through the present analysis determines
the branching ratio
\begin {equation}
BR [f_0(1500) \to \eta \eta ]/BR [f_0(1500) \to \pi ^0 \pi ^0] =
0.52 \pm 0.16.
\end {equation}
With this branching ratio, the magnitudes of $f_0(1500)\pi$ amplitudes
fitted to $3\pi ^0$ agree naturally for all partial waves
with those fitted to $\eta \eta \pi ^0$.
In the final fit, values of $\Lambda _0$ are set to magnitudes predicted from
$\eta \eta \pi ^0$ and equn. (4).

Contributions from $f_0(1770)$, $f_2(1980)$ and $f_0(2105)$ cannot be
determined reliably from $3\pi ^0$ data, because of the intrinsic
systematic uncertainty in how to treat the broad background.
Branching ratios for $f_0(2105)$ and $f_2(1980)$ between $\eta \eta$
and $\pi ^0 \pi ^0$ have been determined well in Ref. [4] from a
combined fit to $\bar pp \to \eta \eta$ and $\pi ^0 \pi ^0$.
These branching ratios are used to determine $\Lambda _0$ in the
penalty functions from $\eta \eta \pi ^0$ data.
The branching fraction of $f_0(1770)$ to $\pi ^0 \pi ^0$ is fitted freely.
Masses and widths of $\pi (2070)$
and $\pi (2360)$ are fixed at the more reliable values fitted to
$\eta \eta \pi ^0$ data; free fits to $3\pi ^0$ data are consistent
with those values but with larger errors of $\pm 60$ MeV.

We begin the physics discussion with $\pi ^0 \eta$ and $\pi ^0 \eta '$.
In Ref. [2], two solutions were found.
That remains the case now.
They are shown for $\pi ^0 \eta$ in Fig. 4.
These solutions are quite distinct.
There is no smooth transition from one solution to the other.
Instead, it is necessary to change the signs of coupling constants
of at least two resonances in order to jump from one solution to the
other.
Extensive searches of this type have not located any
further solutions compatible with a simultaneous fit to $3\pi ^0$.
As one varies $r$ parameters, the two previously published solutions
deform continuously to those given here.
The quality of the fits is indistinguishable by eye from those shown in Ref.
[2].
The fit to $\pi ^0 \eta$ and $\pi ^0 \eta '$ is now better
for the new solution 2: $\chi ^2 = 607$ for 432 points, compared with 720
for solution 1.

The essential difference
between these two solutions is that the upper of two $4^+$ resonances
requires $r_4 = 0.87 \pm 0.27$ for solution 2 but $r_4 = 0.30 \pm 0.31$
for solution 1.
This difference is accompanied by large changes of coupling constants
to $\pi ^0 \eta$ and $\pi ^0 \eta '$ for $2^+$ and $4^+$ states.
The $4^+$ intensity is much larger in solution 1 and smaller for $2^+$.

In both solutions a large $0^+$ amplitude is required.
Much of it may be fitted as background from a pole in the region
1450--1750 MeV. However, a resonance at $2005 \pm 30$  MeV with
$\Gamma = 300 \pm 35$ MeV is also required and gives a significant
improvement in $\chi ^2$ of 290.
There is further evidence for this state from $\bar pp \to \eta \pi ^0 \pi ^0
\pi ^0$ data [9].
One expects a further $0^+$ state around 2250--2360 MeV.
However, adding it to either solution 1 or 2 changes $\chi ^2$ by $< 73$,
which is not enough to establish the presence of another state.
The addition of this second $0^+$ state increases the errors for the
lower one.
In the final fit, we therefore fix the mass and width of the lower
state to $M = 2025$ MeV, $\Gamma = 320$ MeV from Ref. [9].

\begin{table}[htb]
\begin{center}
\begin{tabular}{ccccccc}
\hline
Name & $J^P$ & $M$   & $\Gamma$ & $r$ & $\Delta S $ & $M(I=0)$ \\
 &           & (MeV) & (MeV)    & &  & (MeV)\\\hline
$\pi _4$ & $4^-$  & $2250 \pm 15$  &  $215 \pm 25$ &
& 11108 & $2328 \pm 38$ \\
$a_4$ & $4^+$  & $2255 \pm 40$  &  $330 ^{+110}_{-50}$ & $0.87 \pm 0.27$
& 2455 & $2283 \pm 17$ \\
$a_4$ & $4^+$  & $2005 ^{+25}_{-45}$  &  $180 \pm 30$ & $0.0 \pm 0.2$
& 2447 & $2018 \pm 6$ \\
$a_3$ & $3^+$  & $2275 \pm 35$  &  $350 ^{+100}_{-50}$ &
& 3154 & $2303 \pm 15$ \\
$a_3$ & $3^+$  & $2031 \pm 12$  &  $150 \pm 18$ & & 18410 &
$2048 \pm 8$ \\
$a_2$ & $2^+$  & $2255 \pm 20$  &  $230 \pm 15$ & $-2.13 \pm 0.20$
& 2289 & $2293 \pm 13$ \\
$a_2$ & $2^+$  & $2175 \pm 40$  &  $310 ^{+90}_{-45}$ & $-0.05 \pm 0.31$
& 1059 & $2240 \pm 15$ \\
$a_2$ & $2^+$  & $2030 \pm 20$  &  $205 \pm 30$ & $2.65 \pm 0.56$
& 1308 & $2001 \pm 10$ \\
$\pi _2$ & $2^-$& $2245 \pm 60$  &  $320 ^{+100}_{-40}$ &
& 2298 & $2267 \pm 14$ \\
$\pi _2$ & $2^-$  & $2005 \pm 15$  &  $200 \pm 40$ &
& 1633 & $2030 \pm 15$ \\
$a_1$ & $1^+$  & $2270 ^{+55}_{-40}$  &  $305 ^{+70}_{-35}$ &
& 2571 & $2310 \pm 60$  \\
$\pi$ & $0^-$  & $2360 \pm 25$  &  $300 ^{+100}_{-50}$ &
& 1955 & $2285 \pm 20$ \\
$\pi$ & $0^-$  & $2070 \pm 35$  &  $310 ^{+100}_{-50}$ &
& 1656 & $2010 ^{+35}_{-60}$ \\
$a_0$ & $0^+$  & (2025)  &  (320) &
& 21374 & $2040 \pm 38$ \\\hline
$a_2$ & $2^+$  & $1950 ^{+30}_{-70}$  &  $180 ^{+30}_{-70}$ & $-0.05 \pm 0.30$
& 2638 & $1934 \pm 20$ \\
$\pi _2$ & $2^-$  & $(1880)$  &  (255) &
& 7315 & $1860 \pm 15$\\
$a_1$ & $1^+$  & $1930 ^{+30}_{-70}$  &  $155 \pm 45$ &
& 2609 & $1971 \pm 15$ \\
$a_1$ & $1^+$  & $(1640)$  &  (300) & & 232 \\
$\pi $ & $0^-$  & $(1801)$  &  (210) & & 2402 \\\hline
\end {tabular}
\caption {Masses and widths of fitted resonances for the preferred solution 2.
Values in parentheses
are fixed from other data. Entries in the lower half of the table
are below the available mass range and may not be determined reliably.
Column 6 shows changes $\Delta S$ in log likelihood when each
resonance is removed from the fit to $3\pi ^0$ data and all remaining
parameters are re-optimised.
The last column shows masses for $I = 0$, $C = +1$ resonances from
Ref. [4] for comparison.}
\end{center}
\end{table}

Fits to the $3\pi ^0$ data are almost identical for solutions 1 and 2
for all partial waves except $2^+$, $4^+$ and $6^+$, and even there
the main differences are for $^3H_4$ and $^3H_6$.
The fit to $3\pi ^0$ data again favours solution 2 by 1182 in log likelihood.
We show below in Tables 1 and 2 that removing any resonance from the fit
changes log likelihood $S$ by amounts which are typically 2000.
Removing any small amplitude for $f_0(980)$, $f_0(1300)$, $f_0(1500)$, etc.
introduces changes in log likelihood up to 250.

Table 1 shows masses and widths of resonances in the preferred
solution 2.
These results supercede those of Refs. [1] and [2].
Statistical errors are very small, typically 5 MeV for masses.
Errors in the Tables  cover systematic
variations in a large number of alternative fits (e.g. omitting
$f_0(980)\pi$, $f_0(1500)\pi$ or $f_2(1565)\pi$ final states).
The parameters of the $6^+$ state are set to those of $a_6(2450)$ of the
Particle Data Group, but there is little sensitivity to this choice.
For solution 1, masses and widths of all states except $2^+$ and $4^+$
show changes from solution 2 no larger than statistical errors,
i.e. $\sim 5$ MeV.
Table 2 shows masses and widths for $2^+$ and $4^+$ states in this alternative
solution.

\begin{table}[htp]
\begin{center}
\begin{tabular}{cccccc}
\hline
Name & $J^P$ & $M$   & $\Gamma$ & r & $\Delta S$ \\
      & (MeV) & (MeV)    &  \\\hline
$a_4$ & $4^+$  & $2220 \pm 20$  &  $345 \pm 65$& $0.30 \pm 0.31$ & 2852 \\
$a_4$ & $4^+$  & $2035 \pm 20$  &  $135 \pm 45$& $0.0 \pm 0.2$ & 2063 \\
$a_2$ & $2^+$  & $2235 \pm 35$  &  $200 \pm 25$ & $-1.74 \pm 0.36$ & 3658 \\
$a_2$ & $2^+$  & $2135 \pm 45$  &  $305 ^{+90}_{-45}$ & $-0.56 \pm 0.53$
& 1097 \\\hline
\end {tabular}
\caption {Masses and width of resonances in the alternative solution 1.
The last column shows changes $\Delta S$ in log likelihood when each
resonance is removed from the fit and all remaining parameters are
re-optimised.}
\end{center}
\end{table}

The essential change in going from solution 2 to solution 1 is that the
upper $4^+$ state moves down in mass from 2255 MeV to 2220 MeV.
There is a small increase
in the mass of the lower $4^+$ state from 2005 to 2030 MeV.
What is happening is that the two $4^+$ states, which have similar values of
$r_4$, are tending to merge. We have observed elsewhere that
such merging of resonances of the same $J^P$
tends to give small improvements in log likelihood
through interference effects. Generally it should be regarded
with suspicion.

There are two physics reasons for preferring solution 2.
The first is that the
upper $2^+$ and $4^+$ states are closer to those observed for
$I = 0$, $C = +1$,
shown in the final column of Table 1 for comparison.
Although one state might shift significantly in mass, for example
because of the opening of a nearby threshold, systematic differences of
60--105 MeV between $I = 0$ and $I = 1$ states seems  unlikely.
The second indication is that the fit to $I = 0$, $C = +1$ found a
large positive value of $r_4$ for the upper resonance of $2.7 \pm 0.5$,
closer to the present solution 1. Values of $r_2$ are similar
for $I = 0$ and 1.

Despite differences of detail between solutions 1 and 2,
the general pattern of masses and widths for $I = 0$ and $I = 1$ is
similar. It is surprising that most $I = 1$ masses tend to
lie 20 MeV lower than for $I = 0$.
Since solution 1 tends to drag the masses of the upper $2^+$ and
$4^+$ states down, there is the possibility that this ambiguity is
everywhere having the effect of lowering masses, through correlations
between partial waves.
We have tried increasing all masses for $I = 1$
systematically by 20 MeV and refitting. This does not solve the problem:
log likelihood increases by 820, but when masses are released, they
drift down again to the solution of Table 1. The shift in mass between
$I = 1$ and 0 is
dictated largely by good determinations for $a_3(2031)$ and $a_2(2255)$.

\begin{figure}
\begin{center}
\vskip -47.5mm
\hskip -2.85mm
\epsfig{file=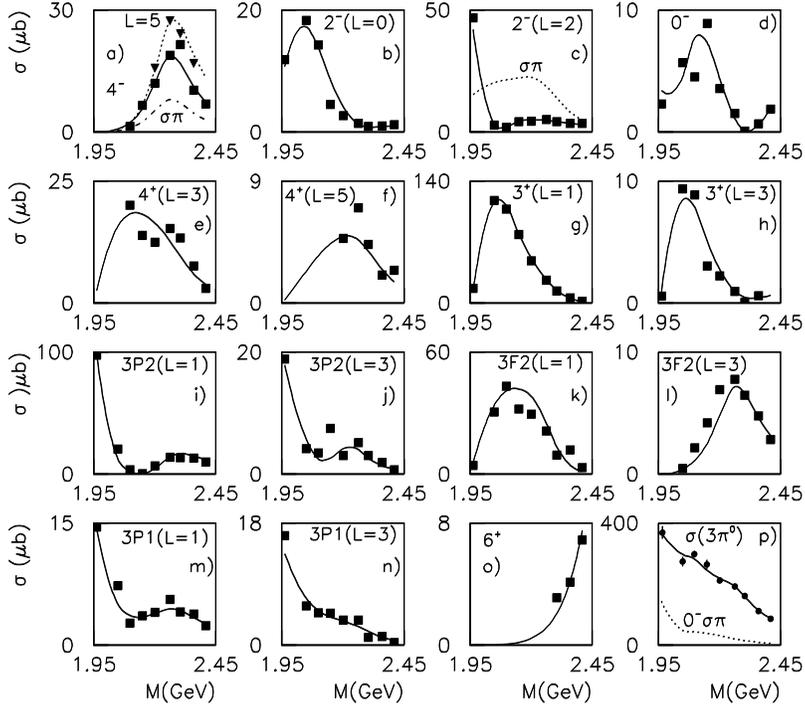,width=13cm}\
\vskip -130.45mm

\epsfig{file=INTENS.EPS,width=13cm}\
~\
\vskip -8mm
\caption {Intensities of individual partial waves (preferred solution 2).
Squares and triangles show free fits at single energies. Full curves are for
$f_2(1270)\pi$ final states. Dotted and chain curves in (a) show respectively
the $4^-$ $f_2\pi$ amplitudes with $L = 5$ (triangles) and the $\sigma \pi$
contribution. Dotted curves in (c) and (p) show $\sigma \pi$ contributions.
In (p), points and the full curve show the integrated cross section for
$3\pi ^0$.}
\end{center}
\end{figure}

Figs. 5 and 6 show intensities and phases of the dominant partial waves
as a function of mass for the preferred solution 2.
The figures also show intensities from fits to individual energies
as squares (or triangles for $4^- [f_2\pi ]_{L=5}$).
In those fits, $r$ values of $2^+$ and $4^+$ states are fixed at each momentum
to values from the full fit. This is the origin of differences from
Figs. 3 and 4 of Ref. [1].
Only magnitudes and phases of $f_2(1270)\pi$ amplitudes are set free in
single-energy fits.
The high partial waves are set free only at
high momenta, where they are well determined.
The scatter of points about the smooth curves indicates the uncertainties,
mostly in phases.

We now comment on individual resonances.
The low mass of the $4^-$ ($^1G_4$) state at 2250 MeV
compared with $\rho _5(2350)$ is surprising
but appears reliable.
There is excellent agreement for this $4^-$ state
with fits to
$\eta \eta \pi ^0$, where it also appears strongly in $a_0(980)\eta$ at
$2255 \pm 30$ MeV with $\Gamma = 185 \pm 60$ MeV.

The $4^+$ states are not accurately defined, because of the lack of
polarisation information to determine values of $r_4$.
For the lower state, $r_4$ is consistent with zero. At this mass,
it would be surprising if $^3H_4$ made any significant contribution,
because of the $\bar pp$ $\ell = 5$ centrifugal barrier.

The lower $3^+$ state at 2031 MeV is  by far the dominant partial
wave at low mass. It is particularly narrow, $\Gamma = 150 \pm 18$ MeV.
This narrow width is essential in order to reproduce the
rapid change in angular distributions shown on Fig. 2 from 600 to 900
MeV/c. The lower $4^+$ state must also be quite narrow, $\Gamma = 180 \pm
30$ MeV (like $f_4(2050)$), for the same reason.
The upper $3^+$ state does not appear as a peak in Fig. 5(g)
because it is overwhelmed by the large amplitude from the lower
resonance. It is required
to explain the phase variation at high mass.
There is also evidence for it from a small peak observed
in the analysis of $\eta \eta \pi ^0$ data [3].

\begin{figure}
\begin{center}
\vskip -47.5mm
\hskip -0.25cm
\epsfig{file=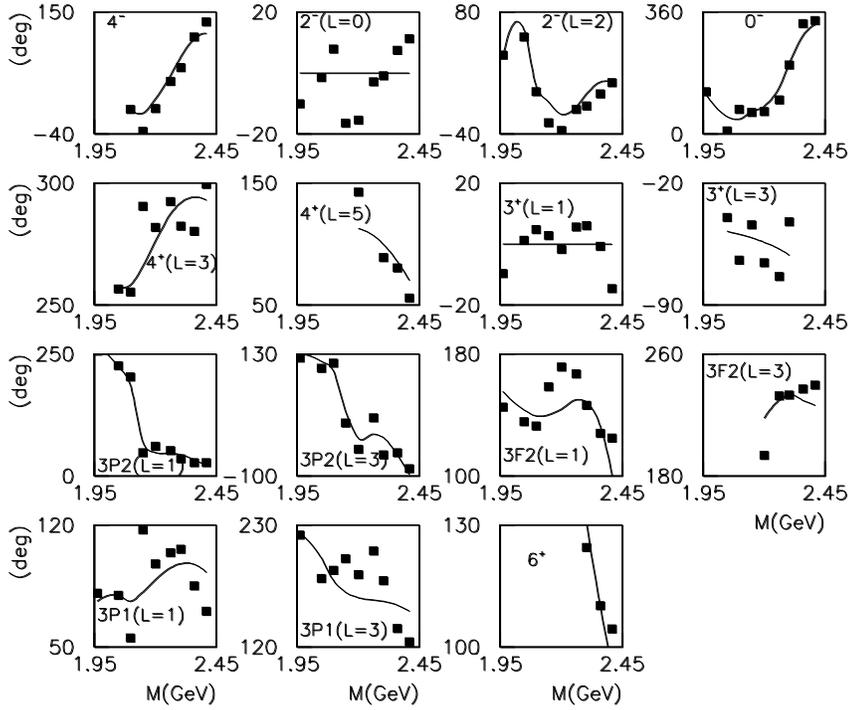,width=13cm}\
\vskip -130.45mm
\epsfig{file=PHASE.EPS,width=13.cm}\
~\
\vskip -8mm
\caption{Phases of individual $f_2\pi$ partial waves,
relative to $2^-(L=0)$ for
singlet states and relative to $3^+$ for triplet partial waves.}
\end{center}
\end{figure}

Two dominantly $^3F_2$ states are required at 2255 and 2030 MeV
and two dominantly $^3P_2$ states at 2175 MeV and $\sim 1950$ MeV.
This pattern is close to that observed for $I = 0$.
The  strong 1950 MeV state is at the bottom
of the available mass range, and its mass and width are strongly
correlated.
If its parameters could be determined accurately elsewhere, this
would stabilise the present analysis considerably.
The $a_2(2175)$ gives the smallest improvement in log likelihood,
namely 1059. It may be simulated to some extent
by changes to mass, width and $r$ value of $a_2(2255)$; its contribution
may also be simulated to a limited extent by a possible contribution from the
missing $0^+$ state in the mass range 2280--2360 MeV.

For $J^P = 2^-$, the behaviour of $L = 2$ and $L = 0$ intensities
of $f_2(1270)\pi$ partial waves at low masses are interesting.
There is a requirement for a very strong $L = 2$ amplitude at the lowest
momentum; it is required specifically to fit the detailed structure of the
mass projection of Fig. 3 at 600 MeV/c. It fits naturally
to $\pi _2(1880)$, reported in an analysis of $\eta \eta \pi ^0 \pi
^0$ data [10]. Despite the fact that this resonance is below
the available mass range, omitting it changes log likelihood by a
particularly large amount, 7315.
A strong $2^- \to [f_2(1270)\pi ]_{L = 2}$ partial wave
in this mass range was reported by Daum et al. [11].
Next, there is a large peak around 2030 MeV in Fig. 5(b) for
$[f_2\pi ]_{L = 0}$.
The rapid variation in these two amplitudes with $L = 0$ and 2 can
be accomodated only by adding a second $\pi _2(2005)$.
Without this extra state, log likelihood is worse by 1633.
There is further evidence for this second $\pi _2$ in the analysis of
$\eta \eta \pi ^0$ data [3].

For $2^-$ at high masses, there is evidence for something around
2245 MeV. Without it, log likelihood is worse by 2298, a
highly significant amount. However, the mass and width are poorly
determined. This is because it couples weakly to $f_2\pi$, and is
observed mostly in decays to $\sigma \pi$, shown by the dotted curve of
Fig. 5(c). Ambiguities in the
treatment of the broad $\sigma$ make the systematic errors for mass
and width large.

The amplitudes for $J^P = 1^+$ are the most difficult to determine,
because of low multiplicity $(2J + 1)$ and because of cross-talk with
$^3P_2$ and $^3F_3$ decays to $f_2\pi$.
There is a definite resonance at $2270$ MeV, Fig. 5(m),
but with  sizeable errors for mass and width. The mass has decreased
somewhat from that reported in the analysis of $3\pi ^0$ data in Ref. [1].
In that previous analysis, a lower $1^+$ state was reported at 2100 MeV.
That claim is now withdrawn.
The addition of the strong $a_2(1950)$ state has improved the fit by a
very large amount and has reduced the $^3P_1$ amplitudes to small values
peaking near threshold. Some low mass $1^+$ contribution is still
required, but it optimises at 1930 MeV, below the available mass range,
which begins at 1960 MeV.
It cannot be regarded as well defined from present data.

\begin{figure}
\begin{center}
\vskip -23mm
\epsfig{file=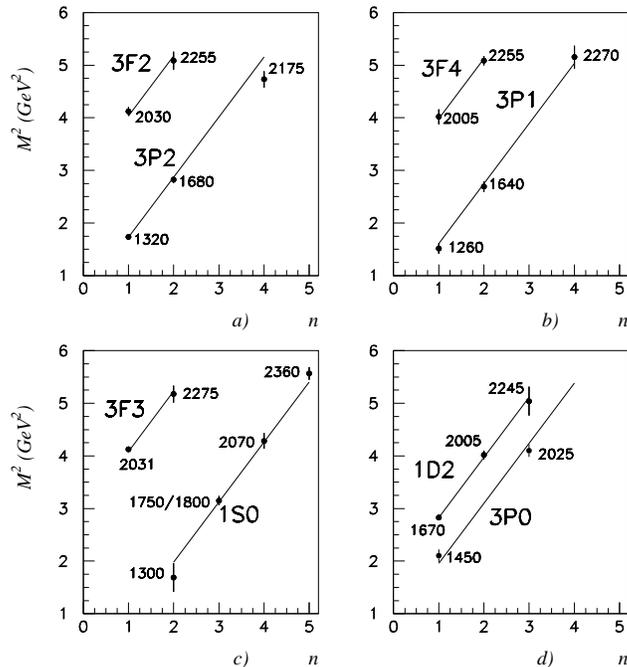,width=10cm}\
~\
\vskip -8mm
\caption{Plots of $M^2$ vs. radial exitation number $n$.
Straight line trajectories are drawn in all cases with a slope of
1.143 GeV$^{-2}$ from Ref. [4]. Numbers indicate masses in MeV.}
\end{center}
\end{figure}

In summary, the mass spectrum for $I = 1$ is similar to that for
$I = 0$.
However, the $\pi _2(2245)$ has large errors;
also $a_2(1950)$ and $a_1(1930)$ are not securely identified in mass and
width, though some such contributions are definitely required.
Fig. 7 shows a comparison of $M^2$ against radial excitation number
with straight line trajectories with a slope of 1.143 GeV$^{-2}$.
This is the average slope fitted to $I = 0$ states in Ref. [4].
The pion is not shown on the $^1S_0$ trajectory, because its mass is
pulled strongly downwards by the instanton interaction.
That may affect $\pi (1300)$ by an unknown amount.
Around 1800 MeV, the VES group reports both $\pi (1800)$ and a $0^-$
peak in $\rho \omega$ at 1750 MeV [12].
If they are distinct, the former is a strong hybrid candidate.
Fig. 7 shows the mean mass with an error covering both possibilities.

Two possible slopes
1.143 GeV$^{-2}$ and 1.38 GeV$^{-2}$ were discussed
in Ref. [13] and trajectories of T-matrix and K-matrix poles were
considered.
The K-matrix poles are not discussed here,
but T-matrix pole trajectories may be investigated fully.
The best solution corresponds much better to the first
slope.
To make an explicit check, we make a set of
fits restricting resonance masses to straight-line trajectories.

That for the $0^{++}$ sector is constructed starting from
$a_0(1450)$.
As discussed above, when two $0^+$ resonances are introduced in
the mass range 1900--2410 MeV, the $0^{++}$ sector becomes weakly
defined and such readjustment causes only very
marginal loss in log likelihood.
The $^3P_2$ trajectory starts from $a_2(1320)$ with
$a_2(1680)$ as the radial excitation.
For a slope of 1.143 GeV$^{-2}$, this requires the mass of the next
state to be 1980-1990 MeV, one standard deviation above
$a_2(1950)$.
The next state is predicted at  $\sim 2250$ MeV.
Although the mass shift required for $a_2(2175)$ is two standard
deviations, we stress that this resonance gives the
smallest contribution to the likelihood value.
The resulting fit with slope 1.143 GeV$^2$ for all
resonances gives log likelihood only 350 worse
than the best fit. There are no visible discrepancies in fits to data
and this solution may be considered acceptable under the restrictions
imposed.
The fit with slope 1.38 GeV$^{-2}$  produces a
solution with log likelihood worse by 2200 than the best solution.
We conclude that resonance masses for the $I=1$ $C=+1$ sector
correspond approximately to the slope of 1.143 GeV$^2$, but not to
the slope 1.38 Gev$^{-2}$ for T-matrix poles.

\section{Acknowledgement}
We thank Prof. V.V. Anisovich for valuable discussions.
We acknowledge financial support from the
British Particle Physics and Astronomy Research Council (PPARC).
The St. Petersburg group wishes to acknowledge financial support from
PPARC and INTAS grant RFBR 95-0267.

\end{document}